\documentclass[twocolumn,english,aps,prb]{revtex4-1}
\usepackage[T1]{fontenc}
\usepackage[latin9]{inputenc}
\usepackage{color}
\usepackage{amsmath}
\usepackage{amssymb}
\usepackage{graphicx}
\usepackage{mhchem}
\usepackage{float}

\makeatletter

\@ifundefined{textcolor}{}
{%
 \definecolor{BLACK}{gray}{0}
 \definecolor{WHITE}{gray}{1}
 \definecolor{RED}{rgb}{1,0,0}
 \definecolor{GREEN}{rgb}{0,1,0}
 \definecolor{BLUE}{rgb}{0,0,1}
 \definecolor{CYAN}{cmyk}{1,0,0,0}
 \definecolor{MAGENTA}{cmyk}{0,1,0,0}
 \definecolor{YELLOW}{cmyk}{0,0,1,0}
}

\usepackage{babel}

\makeatother

\usepackage{babel}
\begin{document}

\title{Chemical analysis of ligand-free silicon nanocrystal surfaces by surface enhanced Raman spectroscopy}

\author{\.{I}lker Do\u{g}an}

\email{i.dogan@tue.nl}

\affiliation{Department of Applied Physics, Eindhoven University of Technology, PO Box 513, 5600 MB Eindhoven, The Netherlands}

\author{Ryan Gresback}

\author{Tomohiro Nozaki}

\affiliation{Department of Mechanical Sciences and Engineering, Tokyo Institute of Technology, 2-12-1, O-okayama, Meguro, 1528550 Tokyo, Japan}

\author{Mauritius C.M. van de Sanden}

\email{m.c.m.vandesanden@differ.nl}

\affiliation{Dutch Institute for Fundamental Energy Research (DIFFER), PO Box 1207, 3430 Nieuwegein, The Netherlands}

\affiliation{Department of Applied Physics, Eindhoven University of Technology, PO Box 513, 5600 MB Eindhoven, The Netherlands}

\begin{abstract}
Surface enhanced Raman spectroscopy (SERS) was used to probe the surface chemistry of chlorine-terminated silicon nanocrystal (Si-NC) surfaces in an air-free environment. SERS effect was observed from the thin films of \ce{Ag_xO} using 514 nm laser wavelength. When a monolayer of Si-NCs were spin-coated on \ce{Ag_xO} SERS substrates, a very clear signal of surface states, including \ce{Si-Cl_x}, and \ce{Si-H_x} were observed. Upon air-exposure, we observed the temporal reduction of \ce{Si-Cl_x} peak intensity, and a development of oxidation-related peak intensities, like \ce{Si-O_x} and \ce{Si-O-H_x}. In addition, first, second and third order transverse optical (TO) modes of Si-NCs were also observed at 519, 1000 and 1600 cm$^{-1}$, respectively. As a comparison, Raman analysis of a thick film (> 200 nm) of Si-NCs deposited on ordinary glass substrates were performed. This analysis only demonstrated the first TO mode of Si-NCs, and the all the other features originated from SERS enhancement did not appear in the spectrum. These results conclude that, SERS is not only capable of single-molecule detection, but also a powerful technique for monitoring the surface chemistry of nanoparticles.
\end{abstract}
\maketitle
Silicon nanocrystals (Si-NCs) have the potential to be an irreplaceable components in the (near)future technological applications by virtue of their size dependent optical, catalytic, and electronic properties. Some of the featured applications, in which Si-NCs play a key role, are light emitting diodes,\cite{Ray2013,Maier-Flaig2013} batteries,\cite{Graetz2003,Liu2005} \ce{CO2}-free fuel production via water splitting,\cite{Erogbogbo2013,Zhang2011} bio-marking,\cite{Erogbogbo2008,Pan2013} and solar cells.\cite{Liu2010,Pi2011,Conibeer2006} Regardless of the nature of the application, surface properties play a critical role on the efficiency, reliability, doping,\cite{Gresback2014,Pereira2014} and compatibility of Si-NCs - as a result of increased surface to volume ratio with respect to bulk Si. Therefore, surface characterization of Si-NCs is of paramount importance.

In order to probe surface chemistry of Si-NCs (and other nanomaterials as well), a surface sensitive technique is necessary. Probing ligand-free Si-NC surfaces and their surface coverage dynamics via oxidation or other means of passivation/functionalization is of interest for optimizing their surface features. Mostly commonly used techniques for chemical analysis of surfaces are x-ray photoelectron spectroscopy (XPS), secondary ion mass spectroscopy (SIMS) and Fourier transform infrared spectroscopy (FTIR). XPS\cite{Dogan2009} and SIMS\cite{Erogbogbo2013} give information on the surface chemistry of materials however, these are destructive techniques as they work by means of removing electrons, and atoms from the surface for analysis of their energies. Therefore, these techniques are unsuitable for probing real-time surface passivation/functionalization dynamics. FTIR is a vibrational spectroscopy, where infrared light passes through a sample and measures the frequency at which energy is absorbed by a specific molecular bond and its various vibrational configurations like stretching or bending modes. FTIR is a non-destructive technique, and gives information on the surface and internal chemistry of the specimen analyzed. Moreover, it is possible to probe the surface chemistry real-time during passivation/functionalization,\cite{Anthony2011,Jariwala2011,Anderson2012} which makes FTIR better suited to XPS and SIMS in this manner.

\begin{figure*}[ht]
  \centering
  \includegraphics[scale=0.45]{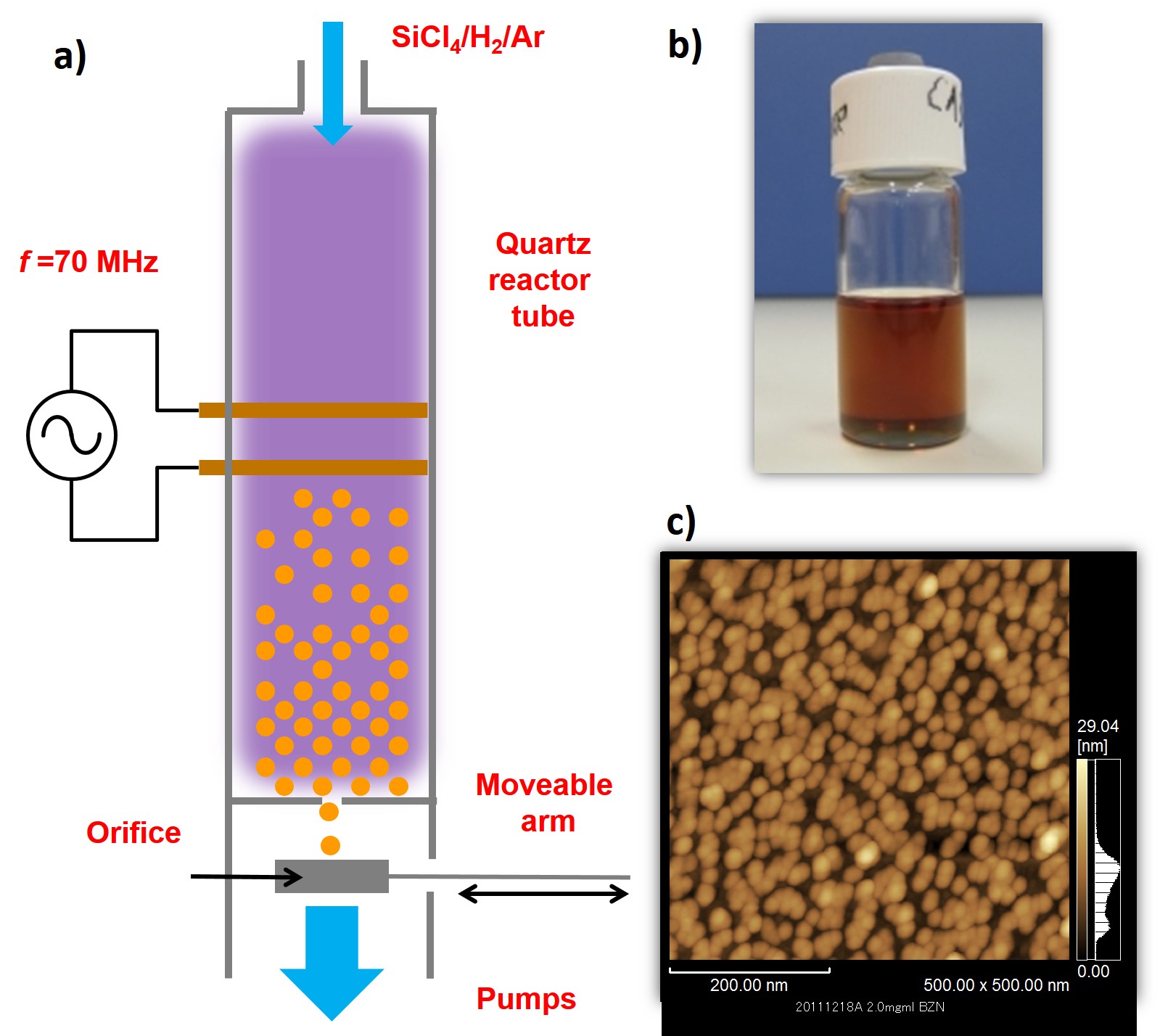}\\
  \caption{(a) A very high frequency (70 MHz) non-thermal plasma is generated using two copper electrodes surrounding a quartz tube, which is employed as a reactor. (b) Si-NCs dispersed in benzonitrile with a density of 1 mg/ml. (c) AFM image of Si-NCs after spin-coating. For this image, Si-wafer was used, however, SERS substrates were coated identically.}
  \label{fig:figure_01}
\end{figure*}

In recent years, Raman spectroscopy, yet another vibrational spectroscopy, has been extensively used in single molecule detection with very high sensitivity due to an effect called surface enhanced Raman scattering (SERS). Surface enhanced Raman scattering (SERS) is widely used for detection of single molecules and low concentration analytes which are impossible to detect by conventional Raman scattering. SERS is observed on special enhancing substrates which show plasmonic effects. Some of the widely used materials can be listed as the thin films or nanostructures of Au,\cite{Zhang2006,Dasary2009} Ag,\cite{Leopold2003,Kneipp1997} and Cu.\cite{Xue1991,Chen2009} In SERS resonances between optical fields and surface plasmons lead to strongly enhanced Raman scattering signals of molecules in the vicinity of metal nanostructures.\cite{Li2012} SERS at extremely high enhancement level brings the effective Raman cross-section to a level of fluorescence cross-section and enables the measurement of Raman spectra from single molecules.\cite{Kneipp1997} The Raman signal from the detected molecules can theoretically be enhanced as high as $10^{14}-10^{15}$ times.\cite{Nie1997} We have previously shown that Raman spectroscopy can be used as a reliable technique to quantitatively analyze the morphology and size distribution of Si-NCs.\cite{Dogan2013} Main advantages of Raman spectroscopy is that, it provides fast, reliable, and non-destructive diagnosis of Si-NCs. In addition to these advantages, as we will show, Raman spectroscopy can also be used to characterize the ligand-free Si-NC surfaces via surface enhanced Raman scattering (SERS).

In this work, we raised the question whether it is possible to monitor the chemical structure of surface molecules of Si-NCs if they are located close to localized surface plasmon resonance (LSPR) zones, which promote SERS effect.  We indeed observed an enhanced Raman scattering signal from a monolayer of Si-NCs spin-coated on an Ag/\ce{Ag_xO} surface under an oxygen-free environment. We also observed that, upon oxidation, we were able to monitor the surface chemistry gradually from partial to full oxidation. In addition, we demonstrated the crystallinity and estimated the size of Si-NCs from the shift in the transverse optical (TO) phonon mode of bulk crystalline silicon. Our findings imply that, Raman spectroscopy can be used as a complete characterization tool for nanomaterials, including the surface analysis by means of SERS.

\begin{figure*}
  \centering
  \includegraphics[scale=0.35]{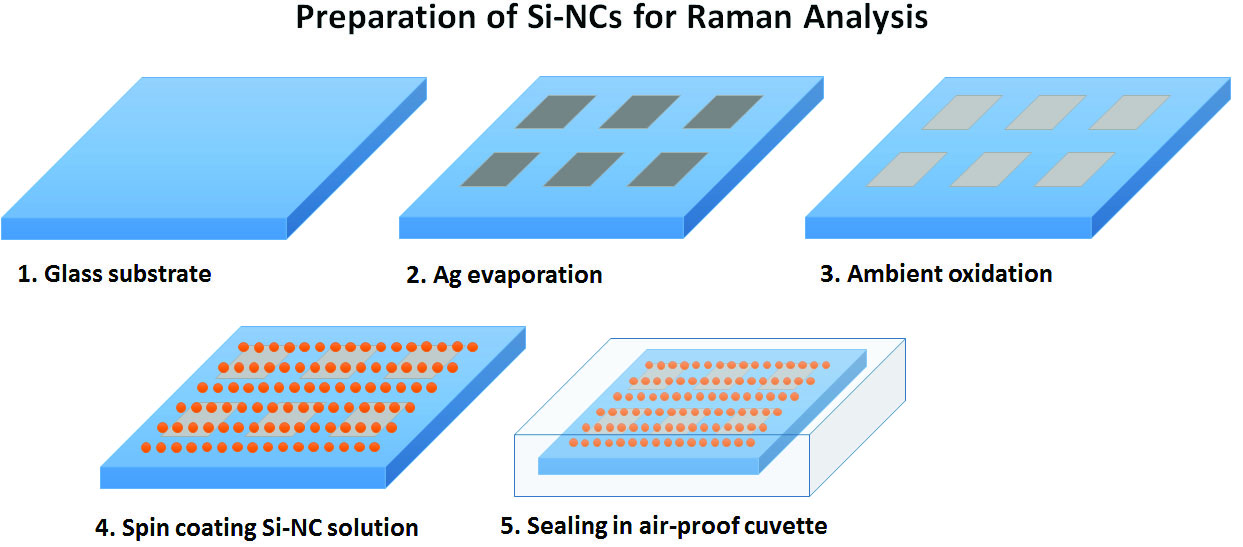}\\
  \caption{Preparation of SERS substrates and spin coating of Si-NC colloidal inks. Step 3 exclusively for \ce{Ag_xO} formation. For Ag and Au deposition, this step is skipped.}
  \label{fig:figure_02}
\end{figure*}

\textbf{Experimental Details}

\emph{a. Synthesis and Solution Processing of Si-NCs}

Free standing silicon nanocrystals were synthesized in a very high frequency (VHF, 70 MHz) non-thermal plasma using a \ce{SiCl4}/\ce{H2}/\ce{Ar} gas mixture as described elsewhere.\cite{Gresback2011} Formation of Si-NCs are favored by dissociation of \ce{SiCl4} molecules through electron impact followed by nucleation and subsequent growth. During the synthesis, Si-NC mean size is controlled by tuning the residence time of the gas flow in the plasma, which resulted diameters of about $\sim$ 5 nm in this work. During the deposition, Si-NCs were collected onto a moveable mesh as illustrated in Figure \ref{fig:figure_01}(a). Nanocrystal collection section of the reactor was decoupled keeping the inside volume in the pressurized Ar atmosphere, and transported into a nitrogen purified glovebox for air-free processing. For the SERS analysis, synthesized Si-NCs were transferred to a nitrogen purified (< 1 ppm oxygen and water) glovebox, without exposure to air. Here, Si-NCs were dispersed and sonicated in anhydrous benzonitrile (Sigma-Aldrich) to form a stable, colloidal nanocrystal ink with chlorine termination. Prepared colloidal ink is diluted to a nanocrystal density of 1 mg/ml (Figure \ref{fig:figure_01}(b)). From this solution, 0.1 ml is drop-casted and spin-coated on SERS substrates with dimensions of 1x2 cm, which resulted in a sub-monolayer of Si-NCs with about 30\% surface coverage according to the atomic force microscopy (AFM) analysis (Figure \ref{fig:figure_01}(c)).

\emph{b. Preparation of SERS Substrates}

Figure \ref{fig:figure_02} shows the preparation sequence of the SERS substrates. In a nitrogen purified glovebox, Au and Ag thin films were deposited through a square shaped mask on  glass substrates by a thermal evaporator. Film thicknesses were monitored by a thickness monitor and were estimated as 10, 50 and 100 nm. Later, Ag coated glass substrates were kept in ambient conditions for 2 min. to allow formation of a top \ce{Ag_xO} (x=2 in the natural oxide) layer. One reference sample was kept air-free to compare the enhancement of Ag and Ag/\ce{Ag_xO}. Following oxidation, Ag/\ce{Ag_xO} samples were processed in a nitrogen glovebox all the samples were drop-casted and spin-coated with Si-NC colloidal ink until all the benzonitrile is dried. To compare the enhancement factor of the SERS substrates, Au, Ag and Ag/\ce{Ag_xO} coated glass substrates were also spin-coated using rhodamine 6G (R6G), a molecule that is known with its high enhancement factor, with a concentration of $10^{-7}$ M in ethanol solution. Finally, R6G and Si-NC spin-coated SERS substrates were put in sealable UV-grade quartz cuvettes for Raman spectroscopy measurements.

\emph{c. Post-analysis Techniques}

Fourier transform infrared spectroscopy (FTIR, JASCO 6100) analyses were conducted by using Si-NCs, which were directly deposited from the plasma onto the metallic mesh grids with enough thickness to collect sufficient signal. The mesh grid with Si-NCs were put inside a vacuum chamber with thallium bromoiodide (KRS-5) window to allow performing measurements in the transmission mode. For Raman spectroscopy measurements, SERS substrates with Si-NCs were used as described above. The Raman laser used had a wavelength of 514 nm \ce{Ar+} ion line. The grid used for the measurements had 1200 lines/mm.

\textbf{Results}

\emph{a. SERS Substrates}

One needs to select carefully the SERS material to be used and the excitation source to perform an effective SERS experiment. The first point to be considered is the type and morphology of the material, and its feasibility to produce the enhancing template in an easy way. The classical SERS materials, Au and Ag are the mostly used ones with different shapes\cite{Li2012,Garcia-Leis2013} and morphologies such as spherically\cite{Urich2012,Felidj2003} shaped nanoparticles, nanowires,\cite{Liao2013,Hunyadi2006} and thin films.\cite{Saito2002} Among these materials and their production techniques, thin film evaporation is simple, and feasible to employ as a SERS template because the production tool, a simple thermal evaporator, is widely available in most of the labs. For this reason, we coated thin films of Au and Ag on glass substrates using a thermal evaporator, which was located in a nitrogen purified glovebox - so that the thin films were not exposed to air. The second point to be considered to perform an effective SERS experiment is the wavelength of the excitation source. It is known that Ag and Au have their unique wavelength ranges in which they support SERS. For Au, this range is in between 600-1250 nm, while for Ag, the SERS support range is in between 400-1000 nm.\cite{Sharma2012} This means that, both Au and Ag are SERS active under red light excitation, while under the infrared light excitation Au is SERS active, and under green light excitation Ag is SERS active.

\begin{figure*}
  \centering
  \includegraphics[scale=0.18]{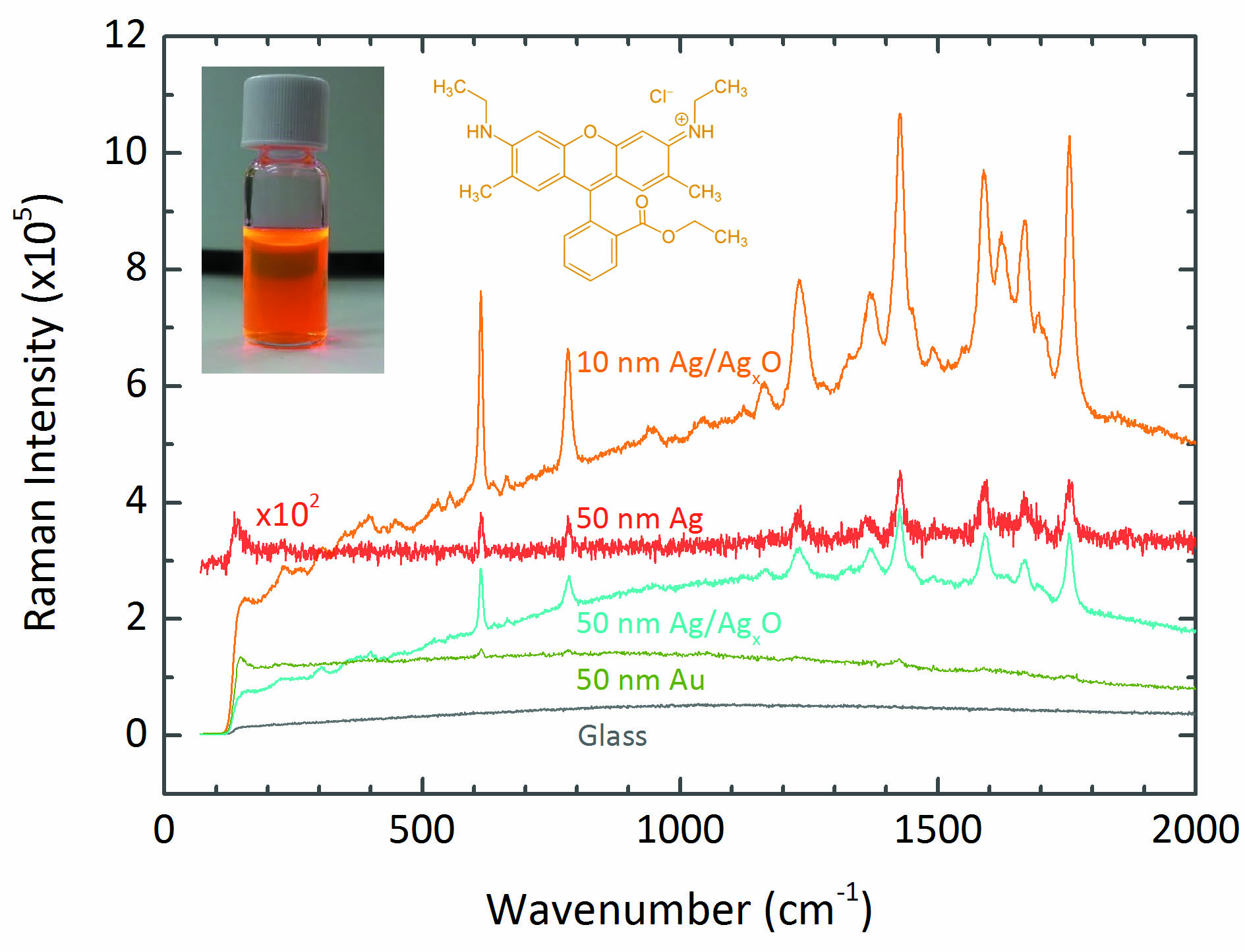}\\
  \caption{Comparison of SERS signals from bare glass and from various thin films evaporated on glass substrates. Highest SERS efficiency was obtained from Ag/\ce{Ag_xO} thin films. }
  \label{fig:figure_03}
\end{figure*}

Figure \ref{fig:figure_03} demonstrates the comparison of SERS signals from $10^{-7}$ M R6G in ethanol, spin-coated on Ag and Au evaporated (50 nm) glass substrates using an excitation wavelength of 514 nm. With respect to the R6G signal from bare glass substrates, which has a complete absence of any spectral features, a clear SERS signal of R6G was observed both from Ag and Au at wavenumbers 614.2, 782.9, 1426.7 and 1753.0 cm$^{-1}$ (for 10 nm and 100 nm Au films, the SERS enhancement was lower with respect to 50 nm). However, as expected, the SERS enhancement was much more evident from Ag thin films using 514 nm light. Observed main R6G vibrational modes are: aromatic C-C stretching modes in the range 1350-1750 cm$^{-1}$, C-C bending mode around 1200 cm$^{-1}$, C-H bending mode around 1150 cm$^{-1}$. Even higher SERS signals with additional features of R6G was observed when the Ag thin films were kept in ambient conditions for 2 minutes before spin-coating with R6G in nitrogen purified atmosphere. In this case, weak C-C stretching modes were also observed at 1140, 1450 and at 1680 cm$^{-1}$, which proves the efficient enhancement from Ag/\ce{Ag_xO} thin films (these features were not observed from R6G molecules spin-coated on Ag thin films). Under ambient conditions, Ag thin film was oxidized, ended up with a shiny white appearance. As a function of film thickness, we found that the highest enhancement was observed from 10 nm Ag/\ce{Ag_xO} films (the enhancement observed from 100 nm films were also similar to that of 50 nm films). The broad feature observed from highly SERS active Ag/\ce{Ag_xO} substrates is due to the fluorescence from R6G (for a visual impression of fluorescence from R6G, see the inset of Figure \ref{fig:figure_03}).

The increase of the enhancement factor upon oxidation of the Ag thin films can be explained in terms of surface modification and improved stability under laser irradiation. Oxidizing the Ag surface induces additional roughness, further leveraging the localized surface plasmon resonance (LSPR) intensities. In addition, \ce{Ag_xO} layers can act as protection barriers for the underlying Ag surface, preventing it from contamination that can possibly come from the adsorbates, and laser irradiation that can lead to agglomeration and particle growth.\cite{Liao2013} Electromagnetic enhancement can be coupled to the molecule to be sensed as long as the distance - the oxide layer thickness - between the LSPR spots and the molecule is within the sensing volume.\cite{Tian2002} The electromagnetic enhancement factor is expressed as the combination of the intensity of local enhancement at the incident frequency and the intensity at the Stokes-shifted frequency, yielding a \textbf{E}$^{4}$ dependency, where \textbf{E} is the electric field.\cite{Stiles2008} The electric field enhancement around a metal particle decays with \emph{d}$^{-3}$, where \emph{d} is the distance to the surface of the particle, \textbf{E}$^{4}$ dependency  of the electromagnetic enhancement leads to a distance dependence of \emph{d}$^{-12}$. Since the surface area scales with \emph{d}$^{2}$ with the size of the LSPR zones get smaller, the overall distance dependency of the SERS intensity is predicted with \emph{d}$^{-10}$ dependence.\cite{Stiles2008,Kennedy1999} In this case, d is the distance between the surface of the SERS active material (Ag thin film) and the probed material (Si-NC surface). Growth of natural \ce{Ag2O} layer on Ag in a pure oxygen at room temperature can go up to 10-20 \AA in an hour,\cite{Rooij1989} but considering the oxidation time of 2 minutes, the oxide layer thickness is probably well below 10 \AA. According to the our observations in Figure \ref{fig:figure_03}, this separation thickness between the Ag surface and Si-NC surfaces is within the SERS enhancement range.

\begin{figure*}
  \centering
  \includegraphics[scale=0.18]{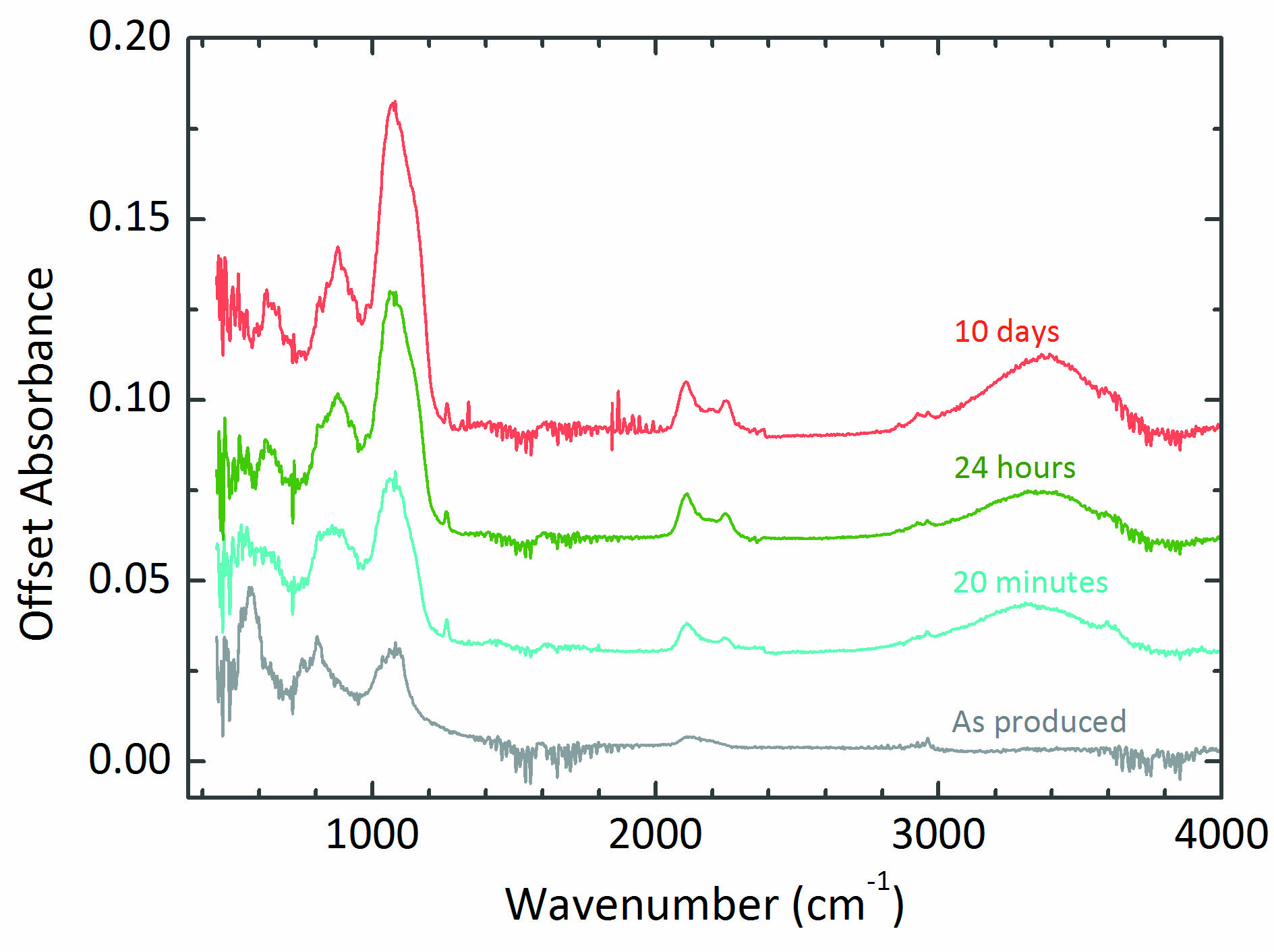}\\
  \caption{FTIR spectra of Si-NCs synthesized in a very high frequency non-thermal plasma. The spectra shows the aging of as-synthesized Si-NCs up to 10 days. Reproduced from ref. 27. \emph{\copyright IOP Publishing. Reproduced with permission. All rights reserved.}}
  \label{fig:figure_04}
\end{figure*}

As a result, we obtained a high SERS activity from 10 nm Ag/\ce{Ag_xO} thin films. Using Ag/\ce{Ag_xO} films is therefore proved as a feasible means of performing SERS experiments since the preparation of Ag thin film is straightforward, a simple evaporation of Ag in a thermal evaporator, and no sophisticated patterning or post-treatment is required. The 10 nm Ag/\ce{Ag_xO} film was used as a SERS substrate for the analysis of Si-NCs.

\emph{b. Analysis of Si-NCs on SERS Substrates}

\emph{Fourier transform infrared spectroscopy.} FTIR was used to probe the surface chemistry of synthesized Si-NCs. Figure \ref{fig:figure_04} demonstrates the as-synthesized Si-NCs and their aging behavior under ambient conditions. In the FTIR spectra we expected to see the surface modes related to hydrogen, chlorine, and oxygen. Hydrogen related modes were observed in the regions 2000-2200 cm$^{-1}$ and 800-950 cm$^{-1}$. These modes are ascribed to \ce{SiH_x} stretching modes and bending modes, respectively. Interestingly, we observed an increase of the \ce{SiH_x} modes intensity upon exposure to air.\cite{Gresback2011} Chlorine related modes were observed with highest intensity from the as-synthesize Si-NCs and their intensities decreased with exposure to air. These modes of \ce{SiCl_x} are located at 575 cm$^{-1}$ region.\cite{Gresback2011} No OH groups were probed from the surfaces of as-synthesized Si-NCs. As expected, the peak intensity of OH groups at 3300 cm$^{-1}$ increased with longer air exposure times. Although we did not expect to see SiOSi modes before exposing Si-NCs to air, we detected these modes from as-synthesized Si-NC surfaces at the range 1000-1150 cm$^{-1}$, which could be due to residual moisture contamination in the glovebox, or etching of the quartz reactor walls during the nanocrystal synthesis.\cite{Gresback2011} The intensity of SiOSi mode increased with longer exposures to air.

\begin{figure*}
  \centering
  \includegraphics[scale=0.135]{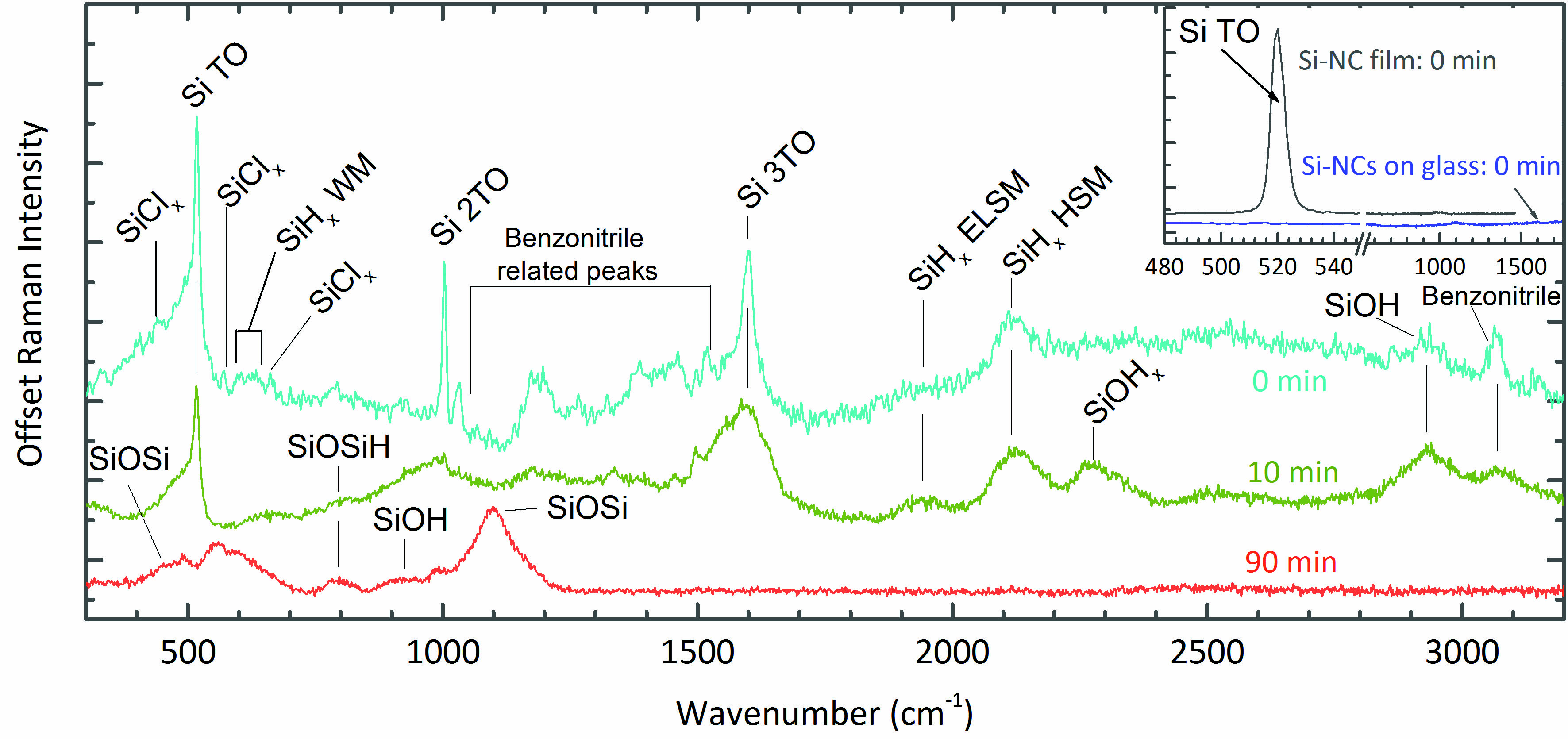}\\
  \caption{SERS analyses of Si-NCs on Ag/\ce{Ag_xO} thin films as a function of air-exposure time: 0 min (non-oxidized, air free analysis), 10 min, and 90 min. Inset shows a measurement from Si-NCs on glass substrates with the same nanocrystal density as SERS analyses (see Figure \ref{fig:figure_02} for more information) and a measurement from thick Si-NC films (thickness >200 nm).}
  \label{fig:figure_05}
\end{figure*}

\emph{Raman spectroscopy.} Raman spectra of Si-NCs spin-coated on Ag/\ce{Ag_xO} SERS substrates are demonstrated in Figure \ref{fig:figure_05}. First, we stress that Si-NCs deposited as a thick film (thickness >200 nm) have a clear Raman signal at $\sim$ 519 cm$^{-1}$ with an estimated size of $\sim$ 4.8 nm from the one particle phonon confinement model.\cite{Dogan2013} The rest of the spectrum does not have any additional features as clearly seen in the inset of Figure \ref{fig:figure_05}. On the other hand, Si-NCs measured from SERS substrate, but from uncoated glass regions demonstrated a flat character in a broad spectral range, concluding that the phonon intensities of Si-NCs are beyond the detection limit of the Raman spectrometer without any enhancement (Si-NCs on glass sample were measured from the regions, which were not coated by Ag/\ce{Ag_xO}, see Figure \ref{fig:figure_02} for more information). However, Si-NCs analyzed on Ag/\ce{Ag_xO} regions show a clear SERS effect, rich of various peaks to be identified. We will elaborate the detected peaks as a function of air exposure time, namely from 0 (non-oxidized, air-free) to 90 minutes.

Unlike the Raman spectrum of Si-NCs on glass regions of the substrate, a monolayer of non-oxidized Si-NCs on Ag/\ce{Ag_xO} regions (0 min) has a very clear footprint of Si-NC TO phonon mode at 519 cm$^{-1}$ proving the enhancement as seen in Figure \ref{fig:figure_05}. In addition, second (2TO) and third (3TO) order transverse optical modes were also detected at $\sim$ 1000 cm$^{-1}$ and at 1600 cm$^{-1}$, respectively, which was not observed from Si-NC thick film (inset of Figure \ref{fig:figure_05}). This observation further demonstrates the enhanced electron-phonon coupling as a result of SERS effect.\cite{Mercaldo2010} Apart from the Raman active modes, the rest of the spectrum has similar features as the FTIR spectrum (note that observed modes are Raman active modes, they were observed in the literature,\cite{Smith1953,Zhaoping1985} and they are possibly very close to the FTIR active modes): additional features observed from the surface of Si-NCs were \ce{SiCl_x} modes around 450, 540, and 650 cm$^{-1}$,\cite{Smith1953,Zhaoping1985} \ce{SiH_x} wagging modes in the region 600-630 cm$^{-1}$,\cite{Zhaoping1985} and \ce{SiH_x} stretching modes in the region 1900-2200 cm$^{-1}$,\cite{Zhaoping1985} as a result of surface termination by chlorine and hydrogen during synthesis. A striking observation is the ability of SERS resolving \ce{SiH_x} stretching modes (1900 cm$^{-1}$ is the extreme low stretching mode (ELSM), and 2100 cm$^{-1}$ is the high stretching mode (HSM)), which cannot be distinguished in a single FTIR measurement as the signal comes both from the surface and from the bulk of Si-NCs.\cite{Smets2008,Smets2007} This observation indicates that only ELSM and HSM of \ce{SiH_x} were present on the nanocrystal surfaces (as the bulk signal is ruled out since the SERS intensity scales with \emph{d}$^{-10}$ with distance from the LSPR region, which makes it unlikely to detect internal chemistry of Si-NCs\cite{Stiles2008}). Additional peaks in the region 1200-1500 cm$^{-1}$ and 3050 cm$^{-1}$ indicated the presence of benzonitrile fragments attached the surface of Si-NCs. An SiOH presence was also detected at 2920 cm$^{-1}$, which is possibly due to the residual water adsorbed on Si-NC surfaces during sample preparation in nitrogen purified glovebox.

After 10 minutes of oxidation of Si-NCs on SERS substrates, a clearly different SERS spectrum was observed with respect to the non-oxidized nanocrystal surfaces. Si-NC TO modes appeared with less intensity, and oxidation-related peaks were appeared: SiOSiH mode at 800 cm$^{-1}$, \ce{SiOH_x} mode at 2250 cm$^{-1}$, and an increased SiOH signal at 2920 cm$^{-1}$. As a result of oxidation, \ce{SiCl_x} and benzonitrile modes disappeared. However, Si-\ce{H_x} modes in the range 1900-2100 cm$^{-1}$ were detected at higher intensities upon oxidation, similar to the FTIR analyses shown in Figure \ref{fig:figure_04}, and to the previous reports on oxidation of chlorine terminated Si-NC surfaces.\cite{Gresback2011,Yasar-Inceoglu2012} After 90 minutes of oxidation, most of the vibrational modes that belong to Si have disappeared. On the other hand, new oxidation-related modes were detected in the range 450-1200 cm$^{-1}$, similar to the FTIR spectra shown in Figure \ref{fig:figure_04}. These are SiOSi modes at 450 and 1100 cm$^{-1}$, SiOSiH mode at 800 cm$^{-1}$ (with an increased intensity with respect to 10 minutes of oxidation), and SiOH stretching mode at 900 cm$^{-1}$.

The reason of vanishing of the main features observed in the SERS spectra previously is possibly due to the increased oxide thickness of Si-NCs (and possibly the further oxidation of Ag thin film), which increases the distance between the LSPR zone and the Si-NC surface beneath the oxide shell, decoupling the plasmonic enhancement. The approximate oxide thickness grown on Si-NCs after 90 minutes of exposure to air is about 1 nm for the originally chlorine terminated Si-NCs according to the Cabrera-Mott mechanism.\cite{Gresback2011,Cabrera1949} Considering the strong distance dependence of SERS intensity (\emph{d}$^{-10}$), the distance between the LSPR zones and Si-NC surfaces in the oxide shell kills the SERS effect, ending up with SERS signal only from the top oxide surface as observed after 90 minutes.

\textbf{Conclusion}

In summary, we have demonstrated the SERS effect from Ag/\ce{Ag_xO} films and used this effect to monitor the surface chemistry of ligand-free Si-NCs. Real time monitoring of nanocrystals surfaces exposed to air revealed the increased oxygen incorporation on nanocrystal surfaces, and the decreased silicon-hydrogen and silicon-chlorine modes, which covered the surface of the as-synthesized nanocrystal surfaces. After 90 minutes of air exposure, the oxide layer on Si-NCs were about 1 nm, which resulted in a spectrum solely containing oxidation related modes. Vanishing of the previously observed features from the non-oxidized Si-NCs was related with the oxide thickness, which resulted in decoupling of the LSPR zones and the nanocrystal surfaces beneath the oxide layer. These observations suggest that SERS can effectively be used for chemical analysis of Si-NC surfaces, which is of critical importance for achieving the desired stability and functionality of Si-NCs for the future technological applications.

\begin{acknowledgments}
This work is part of the research programme of the Foundation for Fundamental Research on Matter (FOM), which is part of the Netherlands Organisation for Scientific Research (NWO). The experimental part of the work was performed in the facilities of Department of Mechanical Sciences and Engineering, Tokyo Institute of Technology. The authors thank R. Yamada and K. Iwazumi for the technical support.
\end{acknowledgments}

\bibliography{SERS}

\begin{thebibliography}{45}%
\makeatletter
\providecommand \@ifxundefined [1]{%
 \@ifx{#1\undefined}
}%
\providecommand \@ifnum [1]{%
 \ifnum #1\expandafter \@firstoftwo
 \else \expandafter \@secondoftwo
 \fi
}%
\providecommand \@ifx [1]{%
 \ifx #1\expandafter \@firstoftwo
 \else \expandafter \@secondoftwo
 \fi
}%
\providecommand \natexlab [1]{#1}%
\providecommand \enquote  [1]{``#1''}%
\providecommand \bibnamefont  [1]{#1}%
\providecommand \bibfnamefont [1]{#1}%
\providecommand \citenamefont [1]{#1}%
\providecommand \href@noop [0]{\@secondoftwo}%
\providecommand \href [0]{\begingroup \@sanitize@url \@href}%
\providecommand \@href[1]{\@@startlink{#1}\@@href}%
\providecommand \@@href[1]{\endgroup#1\@@endlink}%
\providecommand \@sanitize@url [0]{\catcode `\\12\catcode `\$12\catcode
  `\&12\catcode `\#12\catcode `\^12\catcode `\_12\catcode `\%12\relax}%
\providecommand \@@startlink[1]{}%
\providecommand \@@endlink[0]{}%
\providecommand \url  [0]{\begingroup\@sanitize@url \@url }%
\providecommand \@url [1]{\endgroup\@href {#1}{\urlprefix }}%
\providecommand \urlprefix  [0]{URL }%
\providecommand \Eprint [0]{\href }%
\providecommand \doibase [0]{http://dx.doi.org/}%
\providecommand \selectlanguage [0]{\@gobble}%
\providecommand \bibinfo  [0]{\@secondoftwo}%
\providecommand \bibfield  [0]{\@secondoftwo}%
\providecommand \translation [1]{[#1]}%
\providecommand \BibitemOpen [0]{}%
\providecommand \bibitemStop [0]{}%
\providecommand \bibitemNoStop [0]{.\EOS\space}%
\providecommand \EOS [0]{\spacefactor3000\relax}%
\providecommand \BibitemShut  [1]{\csname bibitem#1\endcsname}%
\let\auto@bib@innerbib\@empty
\bibitem [{\citenamefont {Ray}\ \emph {et~al.}(2013)\citenamefont {Ray},
  \citenamefont {Maikap}, \citenamefont {Banerjee},\ and\ \citenamefont
  {Das}}]{Ray2013}%
  \BibitemOpen
  \bibfield  {author} {\bibinfo {author} {\bibfnamefont {S.~K.}\ \bibnamefont
  {Ray}}, \bibinfo {author} {\bibfnamefont {S.}~\bibnamefont {Maikap}},
  \bibinfo {author} {\bibfnamefont {W.}~\bibnamefont {Banerjee}}, \ and\
  \bibinfo {author} {\bibfnamefont {S.}~\bibnamefont {Das}},\ }\href {\doibase
  10.1088/0022-3727/46/15/153001} {\bibfield  {journal} {\bibinfo  {journal}
  {J. Phys. D: Appl. Phys.}\ }\textbf {\bibinfo {volume} {46}},\ \bibinfo
  {pages} {153001} (\bibinfo {year} {2013})}\BibitemShut {NoStop}%
\bibitem [{\citenamefont {Maier-Flaig}\ \emph {et~al.}(2013)\citenamefont
  {Maier-Flaig}, \citenamefont {Rinck}, \citenamefont {Stephan}, \citenamefont
  {Bocksrocker}, \citenamefont {Bruns}, \citenamefont {K\"{u}bel},
  \citenamefont {Powell}, \citenamefont {Ozin},\ and\ \citenamefont
  {Lemmer}}]{Maier-Flaig2013}%
  \BibitemOpen
  \bibfield  {author} {\bibinfo {author} {\bibfnamefont {F.}~\bibnamefont
  {Maier-Flaig}}, \bibinfo {author} {\bibfnamefont {J.}~\bibnamefont {Rinck}},
  \bibinfo {author} {\bibfnamefont {M.}~\bibnamefont {Stephan}}, \bibinfo
  {author} {\bibfnamefont {T.}~\bibnamefont {Bocksrocker}}, \bibinfo {author}
  {\bibfnamefont {M.}~\bibnamefont {Bruns}}, \bibinfo {author} {\bibfnamefont
  {C.}~\bibnamefont {K\"{u}bel}}, \bibinfo {author} {\bibfnamefont {A.~K.}\
  \bibnamefont {Powell}}, \bibinfo {author} {\bibfnamefont {G.~A.}\
  \bibnamefont {Ozin}}, \ and\ \bibinfo {author} {\bibfnamefont
  {U.}~\bibnamefont {Lemmer}},\ }\href {\doibase 10.1021/nl3038689} {\bibfield
  {journal} {\bibinfo  {journal} {Nano Lett.}\ }\textbf {\bibinfo {volume}
  {13}},\ \bibinfo {pages} {1} (\bibinfo {year} {2013})}\BibitemShut {NoStop}%
\bibitem [{\citenamefont {Graetz}\ \emph {et~al.}(2003)\citenamefont {Graetz},
  \citenamefont {Ahn}, \citenamefont {Yazami},\ and\ \citenamefont
  {Fultz}}]{Graetz2003}%
  \BibitemOpen
  \bibfield  {author} {\bibinfo {author} {\bibfnamefont {J.}~\bibnamefont
  {Graetz}}, \bibinfo {author} {\bibfnamefont {C.~C.}\ \bibnamefont {Ahn}},
  \bibinfo {author} {\bibfnamefont {R.}~\bibnamefont {Yazami}}, \ and\ \bibinfo
  {author} {\bibfnamefont {B.}~\bibnamefont {Fultz}},\ }\href {\doibase
  10.1149/1.1596917} {\bibfield  {journal} {\bibinfo  {journal} {Electrochem.
  Solid-State Lett.}\ }\textbf {\bibinfo {volume} {6}},\ \bibinfo {pages} {A194
  } (\bibinfo {year} {2003})}\BibitemShut {NoStop}%
\bibitem [{\citenamefont {Liu}\ \emph {et~al.}(2005)\citenamefont {Liu},
  \citenamefont {Guo}, \citenamefont {Young}, \citenamefont {Shieh},
  \citenamefont {Wu}, \citenamefont {Yang},\ and\ \citenamefont
  {Wu}}]{Liu2005}%
  \BibitemOpen
  \bibfield  {author} {\bibinfo {author} {\bibfnamefont {W.-R.}\ \bibnamefont
  {Liu}}, \bibinfo {author} {\bibfnamefont {Z.-Z.}\ \bibnamefont {Guo}},
  \bibinfo {author} {\bibfnamefont {W.-S.}\ \bibnamefont {Young}}, \bibinfo
  {author} {\bibfnamefont {D.-T.}\ \bibnamefont {Shieh}}, \bibinfo {author}
  {\bibfnamefont {H.-C.}\ \bibnamefont {Wu}}, \bibinfo {author} {\bibfnamefont
  {M.-H.}\ \bibnamefont {Yang}}, \ and\ \bibinfo {author} {\bibfnamefont
  {N.-L.}\ \bibnamefont {Wu}},\ }\href {\doibase
  10.1016/j.jpowsour.2004.07.032} {\bibfield  {journal} {\bibinfo  {journal}
  {J. Power Sources}\ }\textbf {\bibinfo {volume} {140}},\ \bibinfo {pages}
  {139} (\bibinfo {year} {2005})}\BibitemShut {NoStop}%
\bibitem [{\citenamefont {Erogbogbo}\ \emph {et~al.}(2013)\citenamefont
  {Erogbogbo}, \citenamefont {Lin}, \citenamefont {Tucciarone}, \citenamefont
  {Lajoie}, \citenamefont {Lai}, \citenamefont {Patki}, \citenamefont
  {Prasad},\ and\ \citenamefont {Swihart}}]{Erogbogbo2013}%
  \BibitemOpen
  \bibfield  {author} {\bibinfo {author} {\bibfnamefont {F.}~\bibnamefont
  {Erogbogbo}}, \bibinfo {author} {\bibfnamefont {T.}~\bibnamefont {Lin}},
  \bibinfo {author} {\bibfnamefont {P.~M.}\ \bibnamefont {Tucciarone}},
  \bibinfo {author} {\bibfnamefont {K.~M.}\ \bibnamefont {Lajoie}}, \bibinfo
  {author} {\bibfnamefont {L.}~\bibnamefont {Lai}}, \bibinfo {author}
  {\bibfnamefont {G.~D.}\ \bibnamefont {Patki}}, \bibinfo {author}
  {\bibfnamefont {P.~N.}\ \bibnamefont {Prasad}}, \ and\ \bibinfo {author}
  {\bibfnamefont {M.~T.}\ \bibnamefont {Swihart}},\ }\href {\doibase
  10.1021/nl304680w} {\bibfield  {journal} {\bibinfo  {journal} {Nano Lett.}\
  }\textbf {\bibinfo {volume} {12}},\ \bibinfo {pages} {451} (\bibinfo {year}
  {2013})}\BibitemShut {NoStop}%
\bibitem [{\citenamefont {Zhang}\ \emph {et~al.}(2011)\citenamefont {Zhang},
  \citenamefont {Liu}, \citenamefont {Wen},\ and\ \citenamefont
  {Jiang}}]{Zhang2011}%
  \BibitemOpen
  \bibfield  {author} {\bibinfo {author} {\bibfnamefont {R.}~\bibnamefont
  {Zhang}}, \bibinfo {author} {\bibfnamefont {X.}~\bibnamefont {Liu}}, \bibinfo
  {author} {\bibfnamefont {Z.}~\bibnamefont {Wen}}, \ and\ \bibinfo {author}
  {\bibfnamefont {Q.}~\bibnamefont {Jiang}},\ }\href
  {http://pubs.acs.org/doi/abs/10.1021/jp111182c} {\bibfield  {journal}
  {\bibinfo  {journal} {J. Phys. Chem. C}\ }\textbf {\bibinfo {volume} {115}},\
  \bibinfo {pages} {3425} (\bibinfo {year} {2011})}\BibitemShut {NoStop}%
\bibitem [{\citenamefont {Erogbogbo}\ \emph {et~al.}(2008)\citenamefont
  {Erogbogbo}, \citenamefont {Yong}, \citenamefont {Roy}, \citenamefont {Xu},
  \citenamefont {Prasad},\ and\ \citenamefont {Swihart}}]{Erogbogbo2008}%
  \BibitemOpen
  \bibfield  {author} {\bibinfo {author} {\bibfnamefont {F.}~\bibnamefont
  {Erogbogbo}}, \bibinfo {author} {\bibfnamefont {K.~K.-t.}\ \bibnamefont
  {Yong}}, \bibinfo {author} {\bibfnamefont {I.}~\bibnamefont {Roy}}, \bibinfo
  {author} {\bibfnamefont {G.~G.}\ \bibnamefont {Xu}}, \bibinfo {author}
  {\bibfnamefont {P.~N.}\ \bibnamefont {Prasad}}, \ and\ \bibinfo {author}
  {\bibfnamefont {M.~T.}\ \bibnamefont {Swihart}},\ }\href
  {http://pubs.acs.org/doi/abs/10.1021/nn700319z} {\bibfield  {journal}
  {\bibinfo  {journal} {ACS Nano}\ }\textbf {\bibinfo {volume} {2}},\ \bibinfo
  {pages} {873} (\bibinfo {year} {2008})}\BibitemShut {NoStop}%
\bibitem [{\citenamefont {Pan}\ \emph {et~al.}(2013)\citenamefont {Pan},
  \citenamefont {Barras}, \citenamefont {Boussekey}, \citenamefont {Qu},
  \citenamefont {Addad},\ and\ \citenamefont {Boukherroub}}]{Pan2013}%
  \BibitemOpen
  \bibfield  {author} {\bibinfo {author} {\bibfnamefont {G.-H.~G.}\
  \bibnamefont {Pan}}, \bibinfo {author} {\bibfnamefont {A.}~\bibnamefont
  {Barras}}, \bibinfo {author} {\bibfnamefont {L.}~\bibnamefont {Boussekey}},
  \bibinfo {author} {\bibfnamefont {X.}~\bibnamefont {Qu}}, \bibinfo {author}
  {\bibfnamefont {A.}~\bibnamefont {Addad}}, \ and\ \bibinfo {author}
  {\bibfnamefont {R.}~\bibnamefont {Boukherroub}},\ }\href {\doibase
  10.1021/la4029468} {\bibfield  {journal} {\bibinfo  {journal} {Langmuir}\
  }\textbf {\bibinfo {volume} {29}},\ \bibinfo {pages} {12688} (\bibinfo {year}
  {2013})}\BibitemShut {NoStop}%
\bibitem [{\citenamefont {Liu}\ and\ \citenamefont
  {Kortshagen}(2010)}]{Liu2010}%
  \BibitemOpen
  \bibfield  {author} {\bibinfo {author} {\bibfnamefont {C.~Y.}\ \bibnamefont
  {Liu}}\ and\ \bibinfo {author} {\bibfnamefont {U.~R.}\ \bibnamefont
  {Kortshagen}},\ }\href {\doibase 10.1007/s11671-010-9632-z} {\bibfield
  {journal} {\bibinfo  {journal} {Nanoscale Res. Lett.}\ }\textbf {\bibinfo
  {volume} {5}},\ \bibinfo {pages} {1253} (\bibinfo {year} {2010})}\BibitemShut
  {NoStop}%
\bibitem [{\citenamefont {Pi}\ \emph {et~al.}(2011)\citenamefont {Pi},
  \citenamefont {Li}, \citenamefont {Li},\ and\ \citenamefont {Yang}}]{Pi2011}%
  \BibitemOpen
  \bibfield  {author} {\bibinfo {author} {\bibfnamefont {X.}~\bibnamefont
  {Pi}}, \bibinfo {author} {\bibfnamefont {Q.}~\bibnamefont {Li}}, \bibinfo
  {author} {\bibfnamefont {D.}~\bibnamefont {Li}}, \ and\ \bibinfo {author}
  {\bibfnamefont {D.}~\bibnamefont {Yang}},\ }\href {\doibase
  10.1016/j.solmat.2011.06.010} {\bibfield  {journal} {\bibinfo  {journal}
  {Sol. Energy Mater. Sol. Cells}\ }\textbf {\bibinfo {volume} {95}},\ \bibinfo
  {pages} {2941} (\bibinfo {year} {2011})}\BibitemShut {NoStop}%
\bibitem [{\citenamefont {Conibeer}\ \emph {et~al.}(2006)\citenamefont
  {Conibeer}, \citenamefont {Green}, \citenamefont {Corkish}, \citenamefont
  {Cho}, \citenamefont {Cho}, \citenamefont {Jiang}, \citenamefont
  {Fangsuwannarak}, \citenamefont {Pink}, \citenamefont {Huang}, \citenamefont
  {Puzzer}, \citenamefont {Trupke}, \citenamefont {Richards}, \citenamefont
  {Shalav},\ and\ \citenamefont {Lin}}]{Conibeer2006}%
  \BibitemOpen
  \bibfield  {author} {\bibinfo {author} {\bibfnamefont {G.}~\bibnamefont
  {Conibeer}}, \bibinfo {author} {\bibfnamefont {M.}~\bibnamefont {Green}},
  \bibinfo {author} {\bibfnamefont {R.}~\bibnamefont {Corkish}}, \bibinfo
  {author} {\bibfnamefont {Y.}~\bibnamefont {Cho}}, \bibinfo {author}
  {\bibfnamefont {E.-C.}\ \bibnamefont {Cho}}, \bibinfo {author} {\bibfnamefont
  {C.-W.}\ \bibnamefont {Jiang}}, \bibinfo {author} {\bibfnamefont
  {T.}~\bibnamefont {Fangsuwannarak}}, \bibinfo {author} {\bibfnamefont
  {E.}~\bibnamefont {Pink}}, \bibinfo {author} {\bibfnamefont {Y.}~\bibnamefont
  {Huang}}, \bibinfo {author} {\bibfnamefont {T.}~\bibnamefont {Puzzer}},
  \bibinfo {author} {\bibfnamefont {T.}~\bibnamefont {Trupke}}, \bibinfo
  {author} {\bibfnamefont {B.}~\bibnamefont {Richards}}, \bibinfo {author}
  {\bibfnamefont {A.}~\bibnamefont {Shalav}}, \ and\ \bibinfo {author}
  {\bibfnamefont {K.-l.}\ \bibnamefont {Lin}},\ }\href {\doibase
  10.1016/j.tsf.2005.12.119} {\bibfield  {journal} {\bibinfo  {journal} {Thin
  Solid Films}\ }\textbf {\bibinfo {volume} {511-512}},\ \bibinfo {pages} {654}
  (\bibinfo {year} {2006})}\BibitemShut {NoStop}%
\bibitem [{\citenamefont {Gresback}\ \emph {et~al.}(2014)\citenamefont
  {Gresback}, \citenamefont {Kramer}, \citenamefont {Ding}, \citenamefont
  {Chen}, \citenamefont {Kortshagen},\ and\ \citenamefont
  {Nozaki}}]{Gresback2014}%
  \BibitemOpen
  \bibfield  {author} {\bibinfo {author} {\bibfnamefont {R.}~\bibnamefont
  {Gresback}}, \bibinfo {author} {\bibfnamefont {N.~J.}\ \bibnamefont
  {Kramer}}, \bibinfo {author} {\bibfnamefont {Y.}~\bibnamefont {Ding}},
  \bibinfo {author} {\bibfnamefont {T.}~\bibnamefont {Chen}}, \bibinfo {author}
  {\bibfnamefont {U.~R.}\ \bibnamefont {Kortshagen}}, \ and\ \bibinfo {author}
  {\bibfnamefont {T.}~\bibnamefont {Nozaki}},\ }\href {\doibase
  10.1021/nn500182b} {\bibfield  {journal} {\bibinfo  {journal} {ACS Nano}\
  }\textbf {\bibinfo {volume} {8}},\ \bibinfo {pages} {5650} (\bibinfo {year}
  {2014})}\BibitemShut {NoStop}%
\bibitem [{\citenamefont {Pereira}\ \emph {et~al.}(2014)\citenamefont
  {Pereira}, \citenamefont {Coutinho}, \citenamefont {Niesar}, \citenamefont
  {Oliveira}, \citenamefont {Aigner}, \citenamefont {Wiggers}, \citenamefont
  {Rayson}, \citenamefont {Briddon}, \citenamefont {Brandt},\ and\
  \citenamefont {Stutzmann}}]{Pereira2014}%
  \BibitemOpen
  \bibfield  {author} {\bibinfo {author} {\bibfnamefont {R.~N.}\ \bibnamefont
  {Pereira}}, \bibinfo {author} {\bibfnamefont {J.}~\bibnamefont {Coutinho}},
  \bibinfo {author} {\bibfnamefont {S.}~\bibnamefont {Niesar}}, \bibinfo
  {author} {\bibfnamefont {T.~A.}\ \bibnamefont {Oliveira}}, \bibinfo {author}
  {\bibfnamefont {W.}~\bibnamefont {Aigner}}, \bibinfo {author} {\bibfnamefont
  {H.}~\bibnamefont {Wiggers}}, \bibinfo {author} {\bibfnamefont {M.~J.}\
  \bibnamefont {Rayson}}, \bibinfo {author} {\bibfnamefont {P.~R.}\
  \bibnamefont {Briddon}}, \bibinfo {author} {\bibfnamefont {M.~S.}\
  \bibnamefont {Brandt}}, \ and\ \bibinfo {author} {\bibfnamefont
  {M.}~\bibnamefont {Stutzmann}},\ }\href {\doibase 10.1021/nl500932q}
  {\bibfield  {journal} {\bibinfo  {journal} {Nano Lett.}\ }\textbf {\bibinfo
  {volume} {14}},\ \bibinfo {pages} {3817} (\bibinfo {year}
  {2014})}\BibitemShut {NoStop}%
\bibitem [{\citenamefont {Dogan}\ \emph {et~al.}(2009)\citenamefont {Dogan},
  \citenamefont {Yildiz},\ and\ \citenamefont {Turan}}]{Dogan2009}%
  \BibitemOpen
  \bibfield  {author} {\bibinfo {author} {\bibfnamefont {I.}~\bibnamefont
  {Dogan}}, \bibinfo {author} {\bibfnamefont {I.}~\bibnamefont {Yildiz}}, \
  and\ \bibinfo {author} {\bibfnamefont {R.}~\bibnamefont {Turan}},\ }\href
  {\doibase 10.1016/j.physe.2008.08.036} {\bibfield  {journal} {\bibinfo
  {journal} {Phys. E (Amsterdam, Neth.)}\ }\textbf {\bibinfo {volume} {41}},\
  \bibinfo {pages} {976} (\bibinfo {year} {2009})}\BibitemShut {NoStop}%
\bibitem [{\citenamefont {Anthony}\ \emph {et~al.}(2011)\citenamefont
  {Anthony}, \citenamefont {Rowe}, \citenamefont {Stein}, \citenamefont
  {Yang},\ and\ \citenamefont {Kortshagen}}]{Anthony2011}%
  \BibitemOpen
  \bibfield  {author} {\bibinfo {author} {\bibfnamefont {R.~J.}\ \bibnamefont
  {Anthony}}, \bibinfo {author} {\bibfnamefont {D.~J.}\ \bibnamefont {Rowe}},
  \bibinfo {author} {\bibfnamefont {M.}~\bibnamefont {Stein}}, \bibinfo
  {author} {\bibfnamefont {J.}~\bibnamefont {Yang}}, \ and\ \bibinfo {author}
  {\bibfnamefont {U.}~\bibnamefont {Kortshagen}},\ }\href {\doibase
  10.1002/adfm.201100784} {\bibfield  {journal} {\bibinfo  {journal} {Adv.
  Funct. Mater.}\ }\textbf {\bibinfo {volume} {21}},\ \bibinfo {pages} {4042}
  (\bibinfo {year} {2011})}\BibitemShut {NoStop}%
\bibitem [{\citenamefont {Jariwala}\ \emph {et~al.}(2011)\citenamefont
  {Jariwala}, \citenamefont {Dewey}, \citenamefont {Stradins}, \citenamefont
  {Ciobanu},\ and\ \citenamefont {Agarwal}}]{Jariwala2011}%
  \BibitemOpen
  \bibfield  {author} {\bibinfo {author} {\bibfnamefont {B.~N.}\ \bibnamefont
  {Jariwala}}, \bibinfo {author} {\bibfnamefont {O.~S.}\ \bibnamefont {Dewey}},
  \bibinfo {author} {\bibfnamefont {P.}~\bibnamefont {Stradins}}, \bibinfo
  {author} {\bibfnamefont {C.~V.}\ \bibnamefont {Ciobanu}}, \ and\ \bibinfo
  {author} {\bibfnamefont {S.}~\bibnamefont {Agarwal}},\ }\href {\doibase
  10.1021/am200541p} {\bibfield  {journal} {\bibinfo  {journal} {ACS Appl.
  Mater. Interfaces}\ }\textbf {\bibinfo {volume} {3}},\ \bibinfo {pages}
  {3033} (\bibinfo {year} {2011})}\BibitemShut {NoStop}%
\bibitem [{\citenamefont {Anderson}\ \emph {et~al.}(2012)\citenamefont
  {Anderson}, \citenamefont {Shircliff}, \citenamefont {Macauley},
  \citenamefont {Smith}, \citenamefont {Lee}, \citenamefont {Agarwal},
  \citenamefont {Stradins},\ and\ \citenamefont {Collins}}]{Anderson2012}%
  \BibitemOpen
  \bibfield  {author} {\bibinfo {author} {\bibfnamefont {I.~E.}\ \bibnamefont
  {Anderson}}, \bibinfo {author} {\bibfnamefont {R.~A.}\ \bibnamefont
  {Shircliff}}, \bibinfo {author} {\bibfnamefont {C.}~\bibnamefont {Macauley}},
  \bibinfo {author} {\bibfnamefont {D.~K.}\ \bibnamefont {Smith}}, \bibinfo
  {author} {\bibfnamefont {B.~G.}\ \bibnamefont {Lee}}, \bibinfo {author}
  {\bibfnamefont {S.}~\bibnamefont {Agarwal}}, \bibinfo {author} {\bibfnamefont
  {P.}~\bibnamefont {Stradins}}, \ and\ \bibinfo {author} {\bibfnamefont
  {R.~T.}\ \bibnamefont {Collins}},\ }\href {\doibase 10.1021/jp211569a}
  {\bibfield  {journal} {\bibinfo  {journal} {J. Phys. Chem. C}\ }\textbf
  {\bibinfo {volume} {116}},\ \bibinfo {pages} {3979} (\bibinfo {year}
  {2012})}\BibitemShut {NoStop}%
\bibitem [{\citenamefont {Zhang}\ \emph {et~al.}(2006)\citenamefont {Zhang},
  \citenamefont {Gao}, \citenamefont {Alvarez-Puebla}, \citenamefont {Buriak},\
  and\ \citenamefont {Fenniri}}]{Zhang2006}%
  \BibitemOpen
  \bibfield  {author} {\bibinfo {author} {\bibfnamefont {J.}~\bibnamefont
  {Zhang}}, \bibinfo {author} {\bibfnamefont {Y.}~\bibnamefont {Gao}}, \bibinfo
  {author} {\bibfnamefont {R.~A.}\ \bibnamefont {Alvarez-Puebla}}, \bibinfo
  {author} {\bibfnamefont {J.~M.}\ \bibnamefont {Buriak}}, \ and\ \bibinfo
  {author} {\bibfnamefont {H.}~\bibnamefont {Fenniri}},\ }\href {\doibase
  10.1002/adma.200601368} {\bibfield  {journal} {\bibinfo  {journal} {Adv.
  Mater.}\ }\textbf {\bibinfo {volume} {18}},\ \bibinfo {pages} {3233}
  (\bibinfo {year} {2006})}\BibitemShut {NoStop}%
\bibitem [{\citenamefont {Dasary}\ \emph {et~al.}(2009)\citenamefont {Dasary},
  \citenamefont {Singh}, \citenamefont {Senapati}, \citenamefont {Yu},\ and\
  \citenamefont {Ray}}]{Dasary2009}%
  \BibitemOpen
  \bibfield  {author} {\bibinfo {author} {\bibfnamefont {S.~S.~R.}\
  \bibnamefont {Dasary}}, \bibinfo {author} {\bibfnamefont {A.~K.}\
  \bibnamefont {Singh}}, \bibinfo {author} {\bibfnamefont {D.}~\bibnamefont
  {Senapati}}, \bibinfo {author} {\bibfnamefont {H.}~\bibnamefont {Yu}}, \ and\
  \bibinfo {author} {\bibfnamefont {P.~C.}\ \bibnamefont {Ray}},\ }\href
  {\doibase 10.1021/ja905134d} {\bibfield  {journal} {\bibinfo  {journal} {J.
  Am. Chem. Soc.}\ }\textbf {\bibinfo {volume} {131}},\ \bibinfo {pages}
  {13806} (\bibinfo {year} {2009})}\BibitemShut {NoStop}%
\bibitem [{\citenamefont {Leopold}\ and\ \citenamefont
  {Lendl}(2003)}]{Leopold2003}%
  \BibitemOpen
  \bibfield  {author} {\bibinfo {author} {\bibfnamefont {N.}~\bibnamefont
  {Leopold}}\ and\ \bibinfo {author} {\bibfnamefont {B.}~\bibnamefont
  {Lendl}},\ }\href {\doibase 10.1021/jp027460u} {\bibfield  {journal}
  {\bibinfo  {journal} {J. Phys. Chem. B}\ }\textbf {\bibinfo {volume} {107}},\
  \bibinfo {pages} {5723} (\bibinfo {year} {2003})}\BibitemShut {NoStop}%
\bibitem [{\citenamefont {Kneipp}\ \emph {et~al.}(1997)\citenamefont {Kneipp},
  \citenamefont {Wang}, \citenamefont {Kneipp}, \citenamefont {Perelman},
  \citenamefont {Itzkan}, \citenamefont {Dasari},\ and\ \citenamefont
  {Feld}}]{Kneipp1997}%
  \BibitemOpen
  \bibfield  {author} {\bibinfo {author} {\bibfnamefont {K.}~\bibnamefont
  {Kneipp}}, \bibinfo {author} {\bibfnamefont {Y.}~\bibnamefont {Wang}},
  \bibinfo {author} {\bibfnamefont {H.}~\bibnamefont {Kneipp}}, \bibinfo
  {author} {\bibfnamefont {L.~T.}\ \bibnamefont {Perelman}}, \bibinfo {author}
  {\bibfnamefont {I.}~\bibnamefont {Itzkan}}, \bibinfo {author} {\bibfnamefont
  {R.~R.}\ \bibnamefont {Dasari}}, \ and\ \bibinfo {author} {\bibfnamefont
  {M.~S.}\ \bibnamefont {Feld}},\ }\href {\doibase 10.1103/PhysRevLett.78.1667}
  {\bibfield  {journal} {\bibinfo  {journal} {Phys. Rev. Lett.}\ }\textbf
  {\bibinfo {volume} {78}},\ \bibinfo {pages} {1667} (\bibinfo {year}
  {1997})}\BibitemShut {NoStop}%
\bibitem [{\citenamefont {Xue}\ \emph {et~al.}(1991)\citenamefont {Xue},
  \citenamefont {Ding}, \citenamefont {Lu},\ and\ \citenamefont
  {Dong}}]{Xue1991}%
  \BibitemOpen
  \bibfield  {author} {\bibinfo {author} {\bibfnamefont {G.}~\bibnamefont
  {Xue}}, \bibinfo {author} {\bibfnamefont {J.}~\bibnamefont {Ding}}, \bibinfo
  {author} {\bibfnamefont {P.}~\bibnamefont {Lu}}, \ and\ \bibinfo {author}
  {\bibfnamefont {J.}~\bibnamefont {Dong}},\ }\href {\doibase
  10.1021/j100172a050} {\bibfield  {journal} {\bibinfo  {journal} {J. Phys.
  Chem.}\ }\textbf {\bibinfo {volume} {95}},\ \bibinfo {pages} {7380} (\bibinfo
  {year} {1991})}\BibitemShut {NoStop}%
\bibitem [{\citenamefont {Chen}\ \emph {et~al.}(2009)\citenamefont {Chen},
  \citenamefont {Yu}, \citenamefont {Fujita},\ and\ \citenamefont
  {Chen}}]{Chen2009}%
  \BibitemOpen
  \bibfield  {author} {\bibinfo {author} {\bibfnamefont {L.-Y.}\ \bibnamefont
  {Chen}}, \bibinfo {author} {\bibfnamefont {J.-S.}\ \bibnamefont {Yu}},
  \bibinfo {author} {\bibfnamefont {T.}~\bibnamefont {Fujita}}, \ and\ \bibinfo
  {author} {\bibfnamefont {M.-W.}\ \bibnamefont {Chen}},\ }\href {\doibase
  10.1002/adfm.200801239} {\bibfield  {journal} {\bibinfo  {journal} {Adv.
  Funct. Mater.}\ }\textbf {\bibinfo {volume} {19}},\ \bibinfo {pages} {1221}
  (\bibinfo {year} {2009})}\BibitemShut {NoStop}%
\bibitem [{\citenamefont {Li}\ \emph {et~al.}(2012)\citenamefont {Li},
  \citenamefont {Hu}, \citenamefont {Li}, \citenamefont {Shen}, \citenamefont
  {Xiong}, \citenamefont {Li},\ and\ \citenamefont {Fan}}]{Li2012}%
  \BibitemOpen
  \bibfield  {author} {\bibinfo {author} {\bibfnamefont {X.}~\bibnamefont
  {Li}}, \bibinfo {author} {\bibfnamefont {H.}~\bibnamefont {Hu}}, \bibinfo
  {author} {\bibfnamefont {D.}~\bibnamefont {Li}}, \bibinfo {author}
  {\bibfnamefont {Z.}~\bibnamefont {Shen}}, \bibinfo {author} {\bibfnamefont
  {Q.}~\bibnamefont {Xiong}}, \bibinfo {author} {\bibfnamefont
  {S.}~\bibnamefont {Li}}, \ and\ \bibinfo {author} {\bibfnamefont {H.~J.}\
  \bibnamefont {Fan}},\ }\href {\doibase 10.1021/am300189n} {\bibfield
  {journal} {\bibinfo  {journal} {ACS Appl. Mater. Interfaces}\ }\textbf
  {\bibinfo {volume} {4}},\ \bibinfo {pages} {2180} (\bibinfo {year}
  {2012})}\BibitemShut {NoStop}%
\bibitem [{\citenamefont {Nie}\ and\ \citenamefont {Emory}(1997)}]{Nie1997}%
  \BibitemOpen
  \bibfield  {author} {\bibinfo {author} {\bibfnamefont {S.}~\bibnamefont
  {Nie}}\ and\ \bibinfo {author} {\bibfnamefont {S.~R.}\ \bibnamefont
  {Emory}},\ }\href {\doibase 10.1126/science.275.5303.1102} {\bibfield
  {journal} {\bibinfo  {journal} {Science}\ }\textbf {\bibinfo {volume}
  {275}},\ \bibinfo {pages} {1102} (\bibinfo {year} {1997})}\BibitemShut
  {NoStop}%
\bibitem [{\citenamefont {Dogan}\ and\ \citenamefont {van~de
  Sanden}(2013)}]{Dogan2013}%
  \BibitemOpen
  \bibfield  {author} {\bibinfo {author} {\bibfnamefont {I.}~\bibnamefont
  {Dogan}}\ and\ \bibinfo {author} {\bibfnamefont {M.}~\bibnamefont {van~de
  Sanden}},\ }\href {\doibase 10.1063/1.4824178} {\bibfield  {journal}
  {\bibinfo  {journal} {J. Appl. Phys.}\ }\textbf {\bibinfo {volume} {114}},\
  \bibinfo {pages} {134310} (\bibinfo {year} {2013})}\BibitemShut {NoStop}%
\bibitem [{\citenamefont {Gresback}\ \emph {et~al.}(2011)\citenamefont
  {Gresback}, \citenamefont {Nozaki},\ and\ \citenamefont
  {Okazaki}}]{Gresback2011}%
  \BibitemOpen
  \bibfield  {author} {\bibinfo {author} {\bibfnamefont {R.}~\bibnamefont
  {Gresback}}, \bibinfo {author} {\bibfnamefont {T.}~\bibnamefont {Nozaki}}, \
  and\ \bibinfo {author} {\bibfnamefont {K.}~\bibnamefont {Okazaki}},\ }\href
  {\doibase 10.1088/0957-4484/22/30/305605} {\bibfield  {journal} {\bibinfo
  {journal} {Nanotechnology}\ }\textbf {\bibinfo {volume} {22}},\ \bibinfo
  {pages} {305605} (\bibinfo {year} {2011})}\BibitemShut {NoStop}%
\bibitem [{\citenamefont {Garcia-Leis}\ \emph {et~al.}(2013)\citenamefont
  {Garcia-Leis}, \citenamefont {Garcia-Ramos},\ and\ \citenamefont
  {Sanchez-Cortes}}]{Garcia-Leis2013}%
  \BibitemOpen
  \bibfield  {author} {\bibinfo {author} {\bibfnamefont {A.}~\bibnamefont
  {Garcia-Leis}}, \bibinfo {author} {\bibfnamefont {J.~V.}\ \bibnamefont
  {Garcia-Ramos}}, \ and\ \bibinfo {author} {\bibfnamefont {S.}~\bibnamefont
  {Sanchez-Cortes}},\ }\href {\doibase 10.1021/jp401737y} {\bibfield  {journal}
  {\bibinfo  {journal} {J. Phys. Chem. C}\ }\textbf {\bibinfo {volume} {117}},\
  \bibinfo {pages} {7791} (\bibinfo {year} {2013})}\BibitemShut {NoStop}%
\bibitem [{\citenamefont {Urich}\ \emph {et~al.}(2012)\citenamefont {Urich},
  \citenamefont {Pospischil}, \citenamefont {Furchi}, \citenamefont {Dietze},
  \citenamefont {Unterrainer},\ and\ \citenamefont {Mueller}}]{Urich2012}%
  \BibitemOpen
  \bibfield  {author} {\bibinfo {author} {\bibfnamefont {A.}~\bibnamefont
  {Urich}}, \bibinfo {author} {\bibfnamefont {A.}~\bibnamefont {Pospischil}},
  \bibinfo {author} {\bibfnamefont {M.~M.}\ \bibnamefont {Furchi}}, \bibinfo
  {author} {\bibfnamefont {D.}~\bibnamefont {Dietze}}, \bibinfo {author}
  {\bibfnamefont {K.}~\bibnamefont {Unterrainer}}, \ and\ \bibinfo {author}
  {\bibfnamefont {T.}~\bibnamefont {Mueller}},\ }\href {\doibase
  10.1063/1.4758696} {\bibfield  {journal} {\bibinfo  {journal} {Appl. Phys.
  Lett.}\ }\textbf {\bibinfo {volume} {101}},\ \bibinfo {pages} {153113}
  (\bibinfo {year} {2012})}\BibitemShut {NoStop}%
\bibitem [{\citenamefont {Felidj}\ \emph {et~al.}(2003)\citenamefont {Felidj},
  \citenamefont {Aubard}, \citenamefont {Levi}, \citenamefont {Krenn},
  \citenamefont {Hohenau}, \citenamefont {Schider}, \citenamefont {Leitner},\
  and\ \citenamefont {Aussenegg}}]{Felidj2003}%
  \BibitemOpen
  \bibfield  {author} {\bibinfo {author} {\bibfnamefont {N.}~\bibnamefont
  {Felidj}}, \bibinfo {author} {\bibfnamefont {J.}~\bibnamefont {Aubard}},
  \bibinfo {author} {\bibfnamefont {G.}~\bibnamefont {Levi}}, \bibinfo {author}
  {\bibfnamefont {J.~R.}\ \bibnamefont {Krenn}}, \bibinfo {author}
  {\bibfnamefont {A.}~\bibnamefont {Hohenau}}, \bibinfo {author} {\bibfnamefont
  {G.}~\bibnamefont {Schider}}, \bibinfo {author} {\bibfnamefont
  {A.}~\bibnamefont {Leitner}}, \ and\ \bibinfo {author} {\bibfnamefont
  {F.~R.}\ \bibnamefont {Aussenegg}},\ }\href {\doibase 10.1063/1.1571979}
  {\bibfield  {journal} {\bibinfo  {journal} {Appl. Phys. Lett.}\ }\textbf
  {\bibinfo {volume} {82}},\ \bibinfo {pages} {3095} (\bibinfo {year}
  {2003})}\BibitemShut {NoStop}%
\bibitem [{\citenamefont {Liao}\ \emph {et~al.}(2013)\citenamefont {Liao},
  \citenamefont {Cheng}, \citenamefont {Li}, \citenamefont {Shao},
  \citenamefont {Wang},\ and\ \citenamefont {Lee}}]{Liao2013}%
  \BibitemOpen
  \bibfield  {author} {\bibinfo {author} {\bibfnamefont {F.}~\bibnamefont
  {Liao}}, \bibinfo {author} {\bibfnamefont {L.}~\bibnamefont {Cheng}},
  \bibinfo {author} {\bibfnamefont {J.}~\bibnamefont {Li}}, \bibinfo {author}
  {\bibfnamefont {M.}~\bibnamefont {Shao}}, \bibinfo {author} {\bibfnamefont
  {Z.}~\bibnamefont {Wang}}, \ and\ \bibinfo {author} {\bibfnamefont {S.-T.}\
  \bibnamefont {Lee}},\ }\href {\doibase 10.1039/c2tc00761d} {\bibfield
  {journal} {\bibinfo  {journal} {J. Mater. Chem. C}\ }\textbf {\bibinfo
  {volume} {1}},\ \bibinfo {pages} {1628} (\bibinfo {year} {2013})}\BibitemShut
  {NoStop}%
\bibitem [{\citenamefont {Hunyadi}\ and\ \citenamefont
  {Murphy}(2006)}]{Hunyadi2006}%
  \BibitemOpen
  \bibfield  {author} {\bibinfo {author} {\bibfnamefont {S.~E.}\ \bibnamefont
  {Hunyadi}}\ and\ \bibinfo {author} {\bibfnamefont {C.~J.}\ \bibnamefont
  {Murphy}},\ }\href {\doibase 10.1039/b607116c} {\bibfield  {journal}
  {\bibinfo  {journal} {J. Mater. Chem.}\ }\textbf {\bibinfo {volume} {16}},\
  \bibinfo {pages} {3929} (\bibinfo {year} {2006})}\BibitemShut {NoStop}%
\bibitem [{\citenamefont {Saito}\ \emph {et~al.}(2002)\citenamefont {Saito},
  \citenamefont {Wang}, \citenamefont {Smith},\ and\ \citenamefont
  {Batchelder}}]{Saito2002}%
  \BibitemOpen
  \bibfield  {author} {\bibinfo {author} {\bibfnamefont {Y.}~\bibnamefont
  {Saito}}, \bibinfo {author} {\bibfnamefont {J.}~\bibnamefont {Wang}},
  \bibinfo {author} {\bibfnamefont {D.}~\bibnamefont {Smith}}, \ and\ \bibinfo
  {author} {\bibfnamefont {D.}~\bibnamefont {Batchelder}},\ }\href
  {http://pubs.acs.org/doi/abs/10.1021/la011554y} {\bibfield  {journal}
  {\bibinfo  {journal} {Langmuir}\ }\textbf {\bibinfo {volume} {18}},\ \bibinfo
  {pages} {2000} (\bibinfo {year} {2002})}\BibitemShut {NoStop}%
\bibitem [{\citenamefont {Sharma}\ \emph {et~al.}(2012)\citenamefont {Sharma},
  \citenamefont {Frontiera}, \citenamefont {Henry}, \citenamefont {Ringe},\
  and\ \citenamefont {{Van Duyne}}}]{Sharma2012}%
  \BibitemOpen
  \bibfield  {author} {\bibinfo {author} {\bibfnamefont {B.}~\bibnamefont
  {Sharma}}, \bibinfo {author} {\bibfnamefont {R.~R.}\ \bibnamefont
  {Frontiera}}, \bibinfo {author} {\bibfnamefont {A.-i.}\ \bibnamefont
  {Henry}}, \bibinfo {author} {\bibfnamefont {E.}~\bibnamefont {Ringe}}, \ and\
  \bibinfo {author} {\bibfnamefont {R.~P.}\ \bibnamefont {{Van Duyne}}},\
  }\href {\doibase 10.1016/S1369-7021(12)70017-2} {\bibfield  {journal}
  {\bibinfo  {journal} {Mater. Today}\ }\textbf {\bibinfo {volume} {15}},\
  \bibinfo {pages} {16} (\bibinfo {year} {2012})}\BibitemShut {NoStop}%
\bibitem [{\citenamefont {Tian}\ \emph {et~al.}(2002)\citenamefont {Tian},
  \citenamefont {Ren},\ and\ \citenamefont {Wu}}]{Tian2002}%
  \BibitemOpen
  \bibfield  {author} {\bibinfo {author} {\bibfnamefont {Z.~Z.-q.}\
  \bibnamefont {Tian}}, \bibinfo {author} {\bibfnamefont {B.}~\bibnamefont
  {Ren}}, \ and\ \bibinfo {author} {\bibfnamefont {D.~D.-y.}\ \bibnamefont
  {Wu}},\ }\href {http://pubs.acs.org/doi/abs/10.1021/jp0257449} {\bibfield
  {journal} {\bibinfo  {journal} {J. Phys. Chem. B}\ }\textbf {\bibinfo
  {volume} {106}},\ \bibinfo {pages} {9463} (\bibinfo {year}
  {2002})}\BibitemShut {NoStop}%
\bibitem [{\citenamefont {Stiles}\ \emph {et~al.}(2008)\citenamefont {Stiles},
  \citenamefont {Dieringer}, \citenamefont {Shah},\ and\ \citenamefont {{Van
  Duyne}}}]{Stiles2008}%
  \BibitemOpen
  \bibfield  {author} {\bibinfo {author} {\bibfnamefont {P.~L.}\ \bibnamefont
  {Stiles}}, \bibinfo {author} {\bibfnamefont {J.~A.}\ \bibnamefont
  {Dieringer}}, \bibinfo {author} {\bibfnamefont {N.~C.}\ \bibnamefont {Shah}},
  \ and\ \bibinfo {author} {\bibfnamefont {R.~P.}\ \bibnamefont {{Van
  Duyne}}},\ }\href {\doibase 10.1146/annurev.anchem.1.031207.112814}
  {\bibfield  {journal} {\bibinfo  {journal} {Annu. Rev. Anal. Chem.}\ }\textbf
  {\bibinfo {volume} {1}},\ \bibinfo {pages} {601} (\bibinfo {year}
  {2008})}\BibitemShut {NoStop}%
\bibitem [{\citenamefont {Kennedy}\ \emph {et~al.}(1999)\citenamefont
  {Kennedy}, \citenamefont {Spaeth}, \citenamefont {Dickey},\ and\
  \citenamefont {Carron}}]{Kennedy1999}%
  \BibitemOpen
  \bibfield  {author} {\bibinfo {author} {\bibfnamefont {B.}~\bibnamefont
  {Kennedy}}, \bibinfo {author} {\bibfnamefont {S.}~\bibnamefont {Spaeth}},
  \bibinfo {author} {\bibfnamefont {M.}~\bibnamefont {Dickey}}, \ and\ \bibinfo
  {author} {\bibfnamefont {K.}~\bibnamefont {Carron}},\ }\href
  {http://pubs.acs.org/doi/abs/10.1021/jp984454i} {\bibfield  {journal}
  {\bibinfo  {journal} {J. Phys. Chem. B}\ }\textbf {\bibinfo {volume} {103}},\
  \bibinfo {pages} {3640} (\bibinfo {year} {1999})}\BibitemShut {NoStop}%
\bibitem [{\citenamefont {Rooij}(1989)}]{Rooij1989}%
  \BibitemOpen
  \bibfield  {author} {\bibinfo {author} {\bibfnamefont {A.~D.}\ \bibnamefont
  {Rooij}},\ }\href {http://esmat.esa.int/atox\_on\_ag.pdf} {\bibfield
  {journal} {\bibinfo  {journal} {EsA Journal}\ }\textbf {\bibinfo {volume}
  {13}},\ \bibinfo {pages} {363} (\bibinfo {year} {1989})}\BibitemShut
  {NoStop}%
\bibitem [{\citenamefont {Mercaldo}\ \emph {et~al.}(2010)\citenamefont
  {Mercaldo}, \citenamefont {Esposito}, \citenamefont {Veneri}, \citenamefont
  {Fameli}, \citenamefont {Mirabella},\ and\ \citenamefont
  {Nicotra}}]{Mercaldo2010}%
  \BibitemOpen
  \bibfield  {author} {\bibinfo {author} {\bibfnamefont {L.~V.}\ \bibnamefont
  {Mercaldo}}, \bibinfo {author} {\bibfnamefont {E.~M.}\ \bibnamefont
  {Esposito}}, \bibinfo {author} {\bibfnamefont {P.~D.}\ \bibnamefont
  {Veneri}}, \bibinfo {author} {\bibfnamefont {G.}~\bibnamefont {Fameli}},
  \bibinfo {author} {\bibfnamefont {S.}~\bibnamefont {Mirabella}}, \ and\
  \bibinfo {author} {\bibfnamefont {G.}~\bibnamefont {Nicotra}},\ }\href
  {\doibase 10.1063/1.3501133} {\bibfield  {journal} {\bibinfo  {journal}
  {Appl. Phys. Lett.}\ }\textbf {\bibinfo {volume} {97}},\ \bibinfo {pages}
  {153112} (\bibinfo {year} {2010})}\BibitemShut {NoStop}%
\bibitem [{\citenamefont {Smith}(1953)}]{Smith1953}%
  \BibitemOpen
  \bibfield  {author} {\bibinfo {author} {\bibfnamefont {A.~L.}\ \bibnamefont
  {Smith}},\ }\href {\doibase 10.1063/1.1698730} {\bibfield  {journal}
  {\bibinfo  {journal} {J. Chem. Phys.}\ }\textbf {\bibinfo {volume} {21}},\
  \bibinfo {pages} {1997} (\bibinfo {year} {1953})}\BibitemShut {NoStop}%
\bibitem [{\citenamefont {Zhaoping}\ \emph {et~al.}(1985)\citenamefont
  {Zhaoping}, \citenamefont {Hexiang}, \citenamefont {Guohua},\ and\
  \citenamefont {Xueshu}}]{Zhaoping1985}%
  \BibitemOpen
  \bibfield  {author} {\bibinfo {author} {\bibfnamefont {W.}~\bibnamefont
  {Zhaoping}}, \bibinfo {author} {\bibfnamefont {H.}~\bibnamefont {Hexiang}},
  \bibinfo {author} {\bibfnamefont {L.}~\bibnamefont {Guohua}}, \ and\ \bibinfo
  {author} {\bibfnamefont {Z.}~\bibnamefont {Xueshu}},\ }\href@noop {}
  {\bibfield  {journal} {\bibinfo  {journal} {Chin. Phys. (Beijing, China)}\
  }\textbf {\bibinfo {volume} {5}},\ \bibinfo {pages} {232} (\bibinfo {year}
  {1985})}\BibitemShut {NoStop}%
\bibitem [{\citenamefont {Smets}\ \emph {et~al.}(2008)\citenamefont {Smets},
  \citenamefont {Matsui},\ and\ \citenamefont {Kondo}}]{Smets2008}%
  \BibitemOpen
  \bibfield  {author} {\bibinfo {author} {\bibfnamefont {A.~H.~M.}\
  \bibnamefont {Smets}}, \bibinfo {author} {\bibfnamefont {T.}~\bibnamefont
  {Matsui}}, \ and\ \bibinfo {author} {\bibfnamefont {M.}~\bibnamefont
  {Kondo}},\ }\href {\doibase 10.1063/1.2837536} {\bibfield  {journal}
  {\bibinfo  {journal} {Appl. Phys. Lett.}\ }\textbf {\bibinfo {volume} {92}},\
  \bibinfo {pages} {033506} (\bibinfo {year} {2008})}\BibitemShut {NoStop}%
\bibitem [{\citenamefont {Smets}\ and\ \citenamefont {van~de
  Sanden}(2007)}]{Smets2007}%
  \BibitemOpen
  \bibfield  {author} {\bibinfo {author} {\bibfnamefont {A.}~\bibnamefont
  {Smets}}\ and\ \bibinfo {author} {\bibfnamefont {M.}~\bibnamefont {van~de
  Sanden}},\ }\href {\doibase 10.1103/PhysRevB.76.073202} {\bibfield  {journal}
  {\bibinfo  {journal} {Phys. Rev. B}\ }\textbf {\bibinfo {volume} {76}},\
  \bibinfo {pages} {073202} (\bibinfo {year} {2007})}\BibitemShut {NoStop}%
\bibitem [{\citenamefont {Yasar-Inceoglu}\ \emph {et~al.}(2012)\citenamefont
  {Yasar-Inceoglu}, \citenamefont {Lopez}, \citenamefont {Farshihagro},\ and\
  \citenamefont {Mangolini}}]{Yasar-Inceoglu2012}%
  \BibitemOpen
  \bibfield  {author} {\bibinfo {author} {\bibfnamefont {O.}~\bibnamefont
  {Yasar-Inceoglu}}, \bibinfo {author} {\bibfnamefont {T.}~\bibnamefont
  {Lopez}}, \bibinfo {author} {\bibfnamefont {E.}~\bibnamefont {Farshihagro}},
  \ and\ \bibinfo {author} {\bibfnamefont {L.}~\bibnamefont {Mangolini}},\
  }\href {\doibase 10.1088/0957-4484/23/25/255604} {\bibfield  {journal}
  {\bibinfo  {journal} {Nanotechnology}\ }\textbf {\bibinfo {volume} {23}},\
  \bibinfo {pages} {255604} (\bibinfo {year} {2012})}\BibitemShut {NoStop}%
\bibitem [{\citenamefont {Cabrera}\ and\ \citenamefont
  {Mott}(1949)}]{Cabrera1949}%
  \BibitemOpen
  \bibfield  {author} {\bibinfo {author} {\bibfnamefont {N.}~\bibnamefont
  {Cabrera}}\ and\ \bibinfo {author} {\bibfnamefont {N.~F.}\ \bibnamefont
  {Mott}},\ }\href {\doibase 10.1088/0034-4885/12/1/308} {\bibfield  {journal}
  {\bibinfo  {journal} {Rep. Prog. Phys.}\ }\textbf {\bibinfo {volume} {12}},\
  \bibinfo {pages} {163} (\bibinfo {year} {1949})}\BibitemShut {NoStop}%
\end{thebibliography}%

\end{document}